\documentclass[fleqn,twoside]{article}
\usepackage{espcrc2}

\usepackage{graphicx}
\usepackage[figuresright]{rotating}

\newcommand{\AmS}{{\protect\the\textfont2
  A\kern-.1667em\lower.5ex\hbox{M}\kern-.125emS}}

\newcommand{\gtrsim}{\mathrel{\hbox{\rlap{\lower.55ex \hbox {$\sim$}}
                   \kern-.3em \raise.4ex \hbox{$>$}}}}
\newcommand{\lesssim}{\mathrel{\hbox{\rlap{\lower.55ex \hbox {$\sim$}}
                   \kern-.3em \raise.4ex \hbox{$<$}}}}

\title{The observers' view of (very) long X-ray bursts: they are super!}

\author{Erik~Kuul\-kers\address[ESA/ESTEC]{ESA/ESTEC, Research and Scientific
       Support Department, Keplerlaan 1, NL-2200 AG Noordwijk, The Netherlands}%
       }
       
\begin{document}

\begin{abstract}
In many X-ray point sources on the sky, the X-ray emission arises because hydrogen and/or helium is
accreted onto a neutron star from a nearby donor star. When this matter settles on the neutron star
surface, it will undergo nuclear fusion. For a large range of physical parameters the fusion is unstable.
The resulting thermo-nuclear explosions last from seconds to minutes. They are observed as short flares
in X-rays and are called `type~I X-ray bursts'. Recently, hours-long X-ray flares have been found
in seven X-ray burst sources with the {\it Beppo\-SAX}/WFC, {\it RXTE}/ASM and {\it RXTE}/PCA. They have 
similar properties to the usual X-ray bursts, except they last for two or three orders of magnitude longer (hence
they are referred to as `superbursts'). This can not be understood in the context of the standard 
nuclear-fusion picture mentioned above. Instead, the superbursts are thought to be related to the unstable
burning of the leftovers from the hydrogen and/or helium fusion. I will discuss the observational
properties of these superbursts. 
\vspace{1pc}
\end{abstract}

\maketitle

\section{Type I X-ray bursts}

\begin{figure}[ht]
\resizebox{7.5cm}{!}{\includegraphics[clip, angle=-90, bb=105 60 583 550]{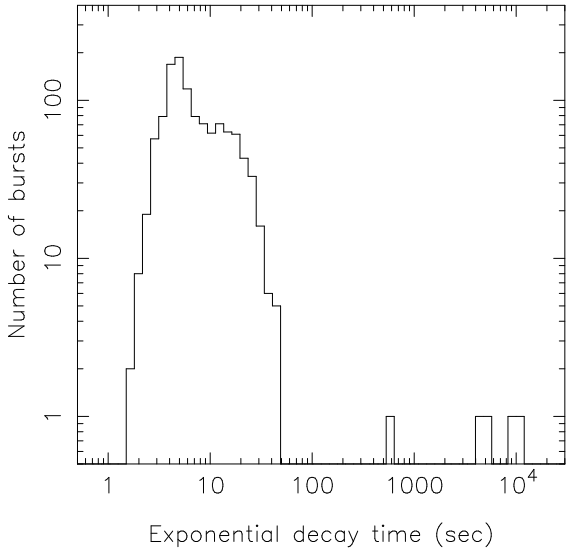}}
\caption{The distribution of the decay times of 1158 X-ray bursts seen by the {\it Beppo\-SAX}/WFCs. 
The decay times are determined from exponential fits to the burst decay profiles. 
Courtesy: the {\it Beppo\-SAX}/WFC team at SRON/Utrecht and CNR/Rome.}
\label{burstdistlog}
\end{figure}

Many low-mass X-ray binaries show thermo-nuclear explosions, or so-called type~I
X-ray bursts (hereafter normal X-ray bursts; for reviews see \cite{Letal1993,SB2003}). 
These appear as rapid ($\sim$1--10\,sec) increases in the X-ray flux, followed by
an exponential-like decline, with typical durations of the order of seconds
to minutes.  They recur with a frequency (typically hours to days) which is 
(partly) set by the supply rate of fresh fuel.
The (net) burst spectra are well described by black-body
emission from a compact object with $\sim$10\,km radius and
inferred temperature of $\sim$1--2\,keV.  The temperature 
increases during the burst rise and decreases
during the decay, reflecting the heating and subsequent cooling of the neutron star surface.
Typical integrated burst energies are 
in the 10$^{39}$ to 10$^{40}$\,erg range.

During some X-ray bursts the energy release is high
enough that the luminosity at the surface of the neutron
star reaches the Eddington limit. At that point the neutron star photosphere expands due to
radiation pressure. Such bursts are referred to as `photospheric radius-expansion type~I X-ray bursts', or
radius-expansion bursts for short. 
During expansion and subsequent contraction the luminosity is expected to
remain almost constant near the Eddington limit. 
Radius expansion bursts are recognized by
an increase in the inferred radius with a simultaneous decrease in the effective temperature near the
peak of an X-ray burst, at approximately constant observed flux.
Note that when the expansion is large the effective
temperature may become so low that the peak of the radiation shifts to UV wavelengths,
and no or little X-rays are emitted. Such events are recognizable by
so-called `precursors' in the X-ray light curves followed by a `main' burst 
\cite{Tetal1984,Letal1984}.

The decay times of X-ray bursts show a bimodal distribution between 
1--50\,sec, with maxima near 5\,sec and 15\,sec (Fig.~\ref{burstdistlog}).
These may be generally attributed to normal X-ray bursts involving either pure He burning or
mixed H/He burning, respectively (see \cite{Letal1993,SB2003}, and 
references therein; see also Cumming, this volume). 

Noticably, Fig.~\ref{burstdistlog} shows 5 events which have long (minutes) 
to very long (hours) decay times. SLX\,1737$-$282 is the source which burst
displayed an exponential decay time of $\sim$10\,min \cite{Zetal2002}.
Note that the X-ray burst seen from
1RXS J171824.2-402934 by the {\it Beppo\-SAX}/WFC has a long decay time as well, i.e., $\gtrsim$200\,sec
\cite{Ketal2000}. Other clear examples of such long X-ray bursts are 
those seen from GX\,17+2 ({\it RXTE}/PCA, \cite{Ketal2002b}) and (possibly) 4U\,1708$-$23 
({\it SAS-3}, \cite{Hetal1978}) which had exponential decay times of $\sim$5\,min.
These long X-ray bursts have durations on the order of half an hour and
energy releases of $\simeq$10$^{41}$\,erg, i.e., typically an order of magnitude
more than normal X-ray bursts. 
The four events with hours long decay times are the subject of this overview,
and are referred to as superbursts. In the next sections I describe the 
phenomenology of these very long X-ray bursts.
I note that the long X-ray bursts discussed above can be accommodated for
in current `normal' X-ray burst theory for those sources accreting at very low
rates ($\lesssim$0.01 times the Eddington accretion rate, see, e.g., \cite{NH2003}).
They do not, however, seem to be related to the superbursts (see, e.g., \cite{Ketal2002b}). 

\section{Superbursts}

The first superburst was discovered by Cornelisse et al.\ to come from the X-ray burster 4U\,1735$-$444 
\cite{Cetal2000}. Another superburst was independently found, originating 
from the X-ray burster 4U\,1820$-$303 \cite{S2000,SB2002}. 
Thereafter, six more events have been seen to occur in five other X-ray bursters 
\cite{W2001,Cetal2002,Ketal2002a,K2002,SM2002,Zetal2003}.
The fact that only eight such events have been found, despite ample observing time, indicates that they
must be rare (see, e.g., \cite{Z2001}).
The recurrence times of these events are, therefore, not well constrained, 
although Wijnands \cite{W2001} reported two superbursts from 4U\,1636$-$536 which 
were $\simeq$4.7~years apart. Observational estimates of the recurrence times are on the 
order of a year (see, e.g., \cite{K2002,Zetal2003}).
The availability of X-ray instruments such as the {\it Beppo\-SAX}/WFC and the {\it RXTE}/ASM, which
more frequently monitor the X-ray sky, and the {\it RXTE}/PCA+HEXTE, which perform
long studies of the X-ray burster population, is
the main reason that superbursts have now been discovered.
In Table~\ref{properties} I show the properties of the superbursts observed so far.

\begin{table*}[ht]
\caption{Properties$^a$ of superbursts ordered along their exponential decay time (after \cite{Ketal2002b})}
\renewcommand{\tabcolsep}{0.2pc} % enlarge column spacing
\begin{tabular}{lccccccc}
\hline
source                                               & {\bf 1820$-$303} & {\bf 1636$-$536} & {\bf Ser~X-1} & {\bf 1735$-$444} & {\bf GX\,3+1}  & {\bf 1731$-$260} & {\bf 1254-690} \\
instrument                                           & PCA      & PCA(ASM) & WFC & WFC      &  ASM & WFC(ASM) & WFC\\
energy range                                         & 2--60\,keV & 2--60\,keV & 2--28\,keV & 2--28\,keV & 2--12\,keV & 2--28\,keV & 2-28\,keV\\
precursor burst?                                     & yes & yes & ? & ? & ? & yes & yes \\
duration (hr)                                        & $>$2.5 & $\sim$6 & $\sim$4 & $\sim$7 & $>$3.3 & $\sim$12 & $\sim$14 \\
$\tau_{\rm rise}$ (min)$^b$                          & $\simeq$2 & $\simeq$14 & $<$45 & $<$36 & $<$117 & $\simeq$20 & $\lesssim$25 \\
$\tau_{\rm exp}$ (hr)                                & $\simeq$1 & 1.05$\pm$0.01 & 1.2$\pm$0.1 & 1.4$\pm$0.1 & 1.6$\pm$0.2 & 2.7$\pm$0.1 & 6.0$\pm$0.3\\
$kT_{\rm max}$ (keV)                                 & $\simeq$3.0 & 2.35$\pm$0.01 & 2.6$\pm$0.2 & 2.6$\pm$0.2 & $\sim$2 & 2.4$\pm$0.1 & 1.8$\pm$0.1 \\
$L_{\rm peak}$ (10$^{38}$\,erg\,sec$^{-1}$)$^{c,d}$  & $\simeq$3.4 & $\simeq$1.3 & $\simeq$1.6 & $\simeq$1.5 & $\sim$0.8 & $\simeq$1.4 & $\simeq$0.4\\
$E_{\rm b}$ (10$^{42}$\,erg)                         & $\gtrsim$1.4 & $\simeq$0.65 & $\simeq$0.8 & $\gtrsim$0.5 & $\gtrsim$0.6 & $\simeq$1.0 & $\simeq$0.8 \\
$\tau$$\equiv$E$_{\rm b}$/L$_{\rm peak}$ (hr)$^d$    & $\gtrsim$1.1 & $\simeq$1.4 & $\simeq$1.4 & $\gtrsim$0.9 & $\gtrsim$2.1 & $\simeq$2.0 & $\simeq$5.0\\
$L_{\rm pers}$ (L$_{\rm Edd}$)$^e$ & $\simeq$0.1     & $\simeq$0.1 & $\simeq$0.2 & $\simeq$0.25 & $\sim$0.2 & $\simeq$0.1 & $\simeq$0.13 \\
$\gamma$$\equiv$L$_{\rm pers}$/L$_{\rm peak}$$^d$    & $\simeq$0.1 & $\simeq$0.3 & $\simeq$0.4 & $\sim$0.4 & $\sim$0.5 & $\simeq$0.4 & $\simeq$0.7\\
$t_{\rm no\,\,bursts}$ (days)$^f$                    & $<$167 & $<$41 & $\sim$34 & $>$7.5 & $<$94 & $>$35 & $<$125\\
H/He or He donor                                     & He & H/He & ? & H/He & ? & ? & H/He\\
references                                           & \cite{S2000,SB2002} & \cite{W2001,SM2002,Ketal2003b} & \cite{Cetal2002} & \cite{Cetal2000} & \cite{K2002} & \cite{Ketal2002a} & \cite{Zetal2003} \\
\hline
\multicolumn{8}{l}{\footnotesize $^a$\,A question mark denotes an unknown value.} \\
\multicolumn{8}{l}{\footnotesize $^b$\,Defined as the time between the peak of the precursor burst and the peak of the superburst.} \\
\multicolumn{8}{l}{\footnotesize $^c$\,Unabsorbed bolometric peak (black-body) luminosity.} \\
\multicolumn{8}{l}{\footnotesize $^d$\,The rise to maximum was seen in 1820$-$303, 1636-536 and 1254-690; values for the others are to be used with caution.} \\
\multicolumn{8}{l}{\footnotesize $^e$\,I used the 0.01--100\,keV unabsorbed flux from spectral fits; the observed maximum flux during radius-expansion bursts} \\
\multicolumn{8}{l}{\footnotesize $^{~}$bursts is used to define the Eddington luminosity.} \\
\multicolumn{8}{l}{\footnotesize $^f$\,Time of cessation of normal X-ray bursts after the superburst.} \\
\end{tabular}
\label{properties}
\end{table*}

\begin{figure}[ht]
\resizebox{7.5cm}{!}{\includegraphics[angle=-90, clip, bb=66 41 585 399]{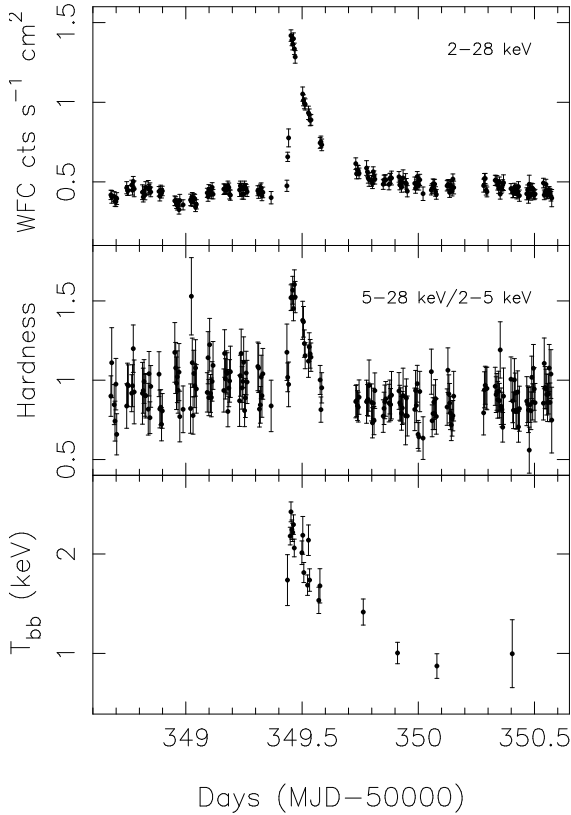}}
\caption{{\it Beppo\-SAX}/WFC light curve (2--28\,keV: {\bf top}) and hardness (ratio of the 
count rates in the 5-28\,keV and 2--5\,keV bands: {\bf middle}) curve showing the 
superburst from KS\,1731$-$260. {\bf bottom:} effective black-body temperature as derived from
black-body model fits to the time-resolved X-ray spectra during the superbursts. The 
pre-burst persistent source X-ray spectrum was subtracted from the spectra during the
superburst. Adapted from \cite{Ketal2002a}.}
\label{ks1731}
\end{figure}

\begin{figure*}[ht]
\resizebox{15.9cm}{!}{\includegraphics[clip, bb=90 285 549 539]{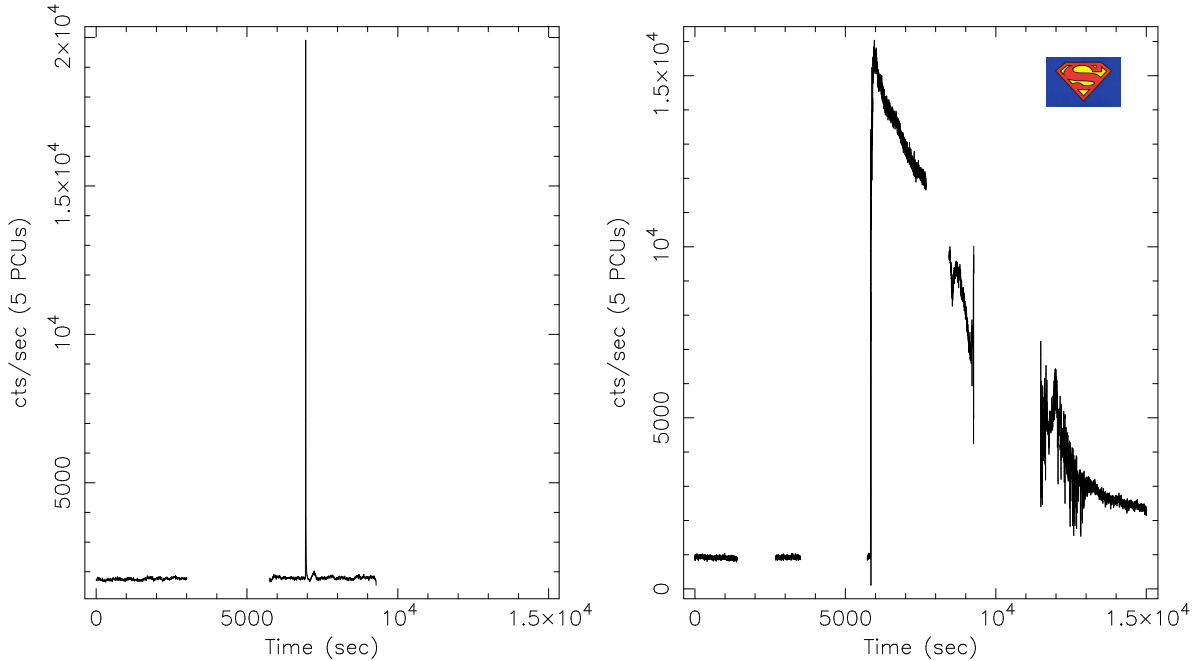}}
\caption{{\it RXTE}/PCA (2--60\,keV) light curves of a normal X-ray burst ({\bf left}) and the 
superburst ({\bf right}) from 4U\,1820$-$30 (see also \cite{SB2002}). 
The time resolution is 0.125\,sec and 1\,sec, respectively. 
The normal X-ray burst and superburst were observed on 1997, May 2 and 1999, Sep 9 (!), respectively.}
\label{1820}
\end{figure*}

In Fig.~\ref{ks1731} (top) I show as an example the superburst from KS\,1731$-$260.
Clearly, the light curve consists of a fast rise and a slower exponential-like decay.
During the rise to superburst maximum the spectrum hardens, whereas during the 
decay the spectrum softens (Fig.~\ref{ks1731}, middle). This is also reflected 
in the spectral fits to the time-resolved pre-burst subtracted X-ray spectra, 
obtained during the superburst (Fig.~\ref{ks1731}, bottom). They are
generally best described by a black-body model (but see below), and the effective
temperature increases and decreases, respectively, during the rise and decay.
These characteristics are similar to those for normal X-ray bursts, and it
was therefore suggested that they are due to thermo-nuclear runaway events
as well \cite{Cetal2000}. 
A big difference with the normal X-ray bursts is the duration, and therefore
the total energy release, of the superburst.
Superbursts last from hours to half a day 
(with exponential decay times of a few hours) and 
integrated fluxes of around 10$^{42}$\,erg (see Table~\ref{properties}).
This is about two or three orders of magnitude more than normal X-ray bursts!
Fig.~\ref{1820} clearly illustrates this difference, by plotting a normal
X-ray burst from 4U\,1820$-$30 and the superburst from 4U\,1820$-$30 on the same time scale.
(The long duration and their enormous energy output 
is the reason they are referred to as `superbursts' \cite{W2001}; see
also the Section `Epilogue' after the reference list.)
 
\begin{figure}[ht]
\resizebox{7.5cm}{!}{\includegraphics[angle=-90, clip, bb=78 48 568 550]{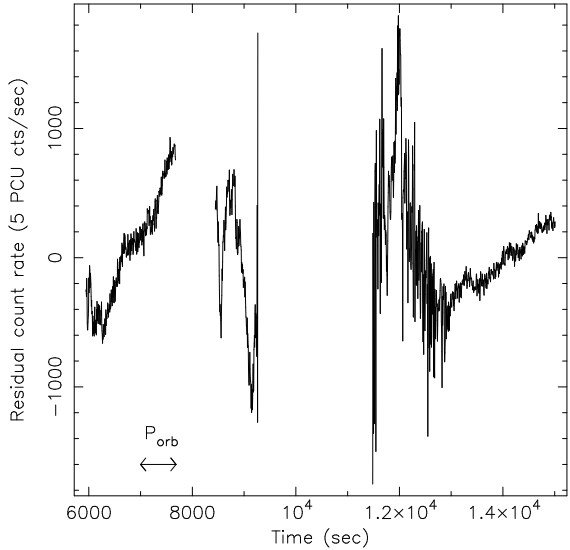}}
\caption{Residual light curve of the superburst from 4U\,1820$-$30 after subtraction
of the exponential fit to the superburst light curve. 
The time resolution is 5\,sec. Indicated in the lower left is the
orbital period (P$_{\rm orb}$$\simeq$685\,sec).}
\label{residual}
\end{figure}

As noted above, the superbursts show exponential-like decays. 
Closer inspection of the superburst in Fig.~\ref{1820} shows, however, that its decay 
exhibits clear deviations from a pure exponential. This is exemplified in 
Fig.~\ref{residual} where I subtracted the exponential fit to the superburst light curve
from the superburst light curve. The corresponding residuals show variations 
on various time scales, but noticably on the orbital period
(11.4\,min \cite{Setal1987}). The time resolved X-ray spectral parameters vary on the same time scale \cite{SB2002}.
A variation is also seen in the superburst light curve of 4U\,1254$-$690,
which is related to its regular X-ray dipping activity \cite{Zetal2003},
but is less clear because of the longer orbital period (3.9\,hr \cite{Cetal1986}).
The superbursts of 4U\,1636$-$536 and 4U\,1735$-$444 were too short (see Table~\ref{properties}) 
to clearly show variations on their orbital periods (3.8 and 4.5\,hr, respectively \cite{Aetal1998}).
No orbital periods are known for Ser\,X-1 and KS\,1731$-$260; their
superburst light curves were consistent with being exponential. The {\it RXTE}/ASM
coverage of the superburst light curve of GX\,3+1 is too sparse to say something meaningful.

\begin{figure*}[ht]
\resizebox{15.9cm}{!}{\includegraphics[angle=-90, clip, bb=191 78 563 770]{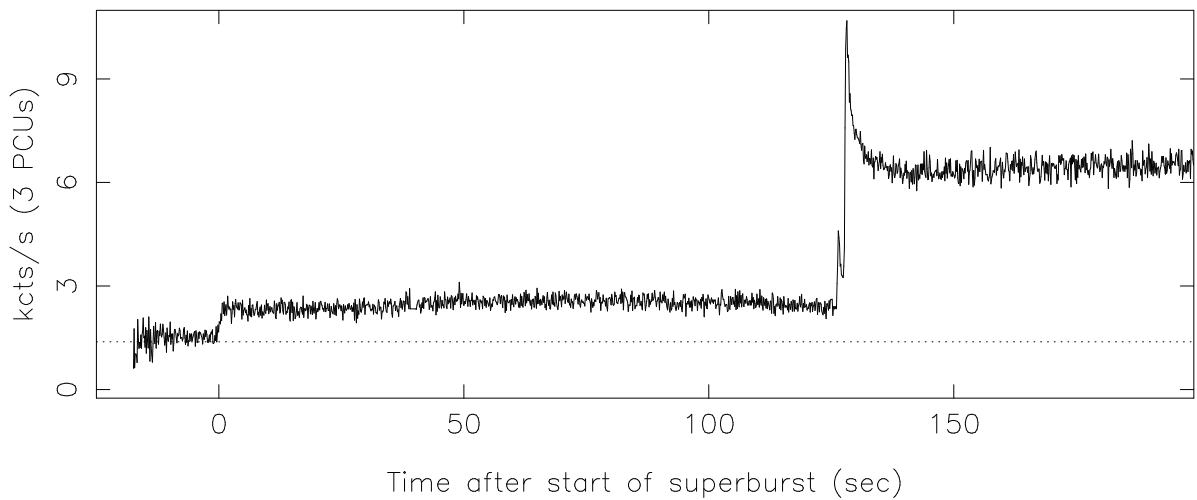}}
\caption{{\it RXTE}/PCA (2--60\,keV) light curve of the start of the superburst from 4U\,1636$-$536
(see also \cite{SM2002}). The time resolution is 0.125\,sec. The first $\sim$70\,sec 
of data were taken during a slew to the source; the light curve has been corrected for background and 
collimator response. The dotted line marks the persistent source flux level in the previous 
{\it RXTE} orbit. Note that the increase near $t=0$\,sec and the subsequent plateau is intrinsic
to the source and not due to the `instrumental' corrections. The strong `variability' 
at the start of the light curve is due to low signal to noise because of low collimator transmissions.}
\label{1636}
\end{figure*}

For four sources observations were available near the start of the superburst.
All of these showed `precursor' normal X-ray bursts near the start of the superburst.
An example is shown in Fig.~\ref{1636}. In 4U\,1636$-$536 the precursor burst showed
up $\sim$125\,sec after the emission had increased suddenly by $\sim$70\%.
The precursor burst is double-peaked, possibly indicating a radius-expansion event. 
This precursor burst is shorter (about 5\,sec) and has a peak flux which is roughly 60\% lower
than normal X-ray bursts from this source
(see, e.g., \cite{S1999}). The shortness of the precursor burst 
and its possible radius-expansion indicates a pure He flash. Immediately
after the precursor burst the superburst had started (see Fig.~\ref{1636}).
Superburst maximum was reached $\simeq$14\,min later.
Similarly, in KS\,1731-260 a weak precursor burst was seen. However, this source
displayed some activity afterwards; the superburst started $\gtrsim$200\,sec 
after the precursor burst. The actual rise to maximum of the superburst was not covered,
but the maximum of the superburst was reached $\simeq$20\,min after the precursor burst.
The precursor burst in 4U\,1254$-$690, on the other hand, was the {\it strongest} 
among previously seen normal X-ray bursts. Like in 4U\,1636$-$536, its superburst 
had started immediately after the precursor burst. 
In 4U\,1636$-$536, KS\,1731$-$260 and 4U\,1254$-$690 the peak flux of the
precursor burst was higher than the superburst peak flux.

The superburst from 4U\,1820$-$303 was immediately preceded by a burst 
(with no recognizable pre-precursor-burst emission like that seen in 4U\,1636$-536$); 
this precursor burst had the same features as its normal X-ray bursts. 
Both the precursor burst and the superburst of 4U\,1820$-$303 were
radius-expansion events; the precursor burst peak flux was somewhat weaker than the superburst peak flux. 
Note that near maximum
of the radius-expansion phase of the superburst the flux dropped even {\it below}
the pre-superburst persistent-source level down to the background-flux level
\cite{SB2002},
indicating the pronounced effect of the superburst on the inner disk regions
where presumably the pre-superburst emission is produced.
The peak of the superburst was reached only $\simeq$2\,min after the precursor burst.
No radius-expansion phase was found during the superburst of 4U\,1636$-$536,
which is consistent with its peak luminosity being lower than peak luminosity reached
during usual radius-expansion bursts (see Table~\ref{properties}).
Similarly, no such phase could be identified in the other superbursts either,
but the rise was not or poorly covered in the other cases.

Analysis of the X-ray spectra during the superburst from 4U\,1820$-$30 revealed the presence of a 
broad emission line between 5.8 and 6.4\,keV, as well as an edge near 8--9\,keV \cite{SB2002}.
Similar deviations from a black-body spectrum appear during the decay part of the 
superburst of 4U\,1636$-$536 \cite{Ketal2003b}. This may be due to reflection of the superburst flux from the inner accretion disk. 
Note that qualitatively similar residuals have been seen during strong normal X-ray bursts
(\cite{Petal1990}; \cite{Ketal2003a}, and references therein; see also \cite{Betal2003}).

Highly coherent pulsations during a superburst of 4U\,1636$-$536 were found near 1.72\,ms
\cite{SM2002}.
The pulsations were detected during an 800\,sec interval near the maximum
of the superburst (note that only during two intervals high time 
resolution data were obtained: $\sim$2500\,sec near the peak of the superburst and 
$\sim$4000\,sec in the decaying tail, see \cite{SM2002}).
Within the 800\,sec interval the frequency increased 
monotonically from 581.89 to 581.93\,Hz, consistent with the predicted orbital motion of the neutron star
around the donor star during this interval. The average pulse
profile was sinusoidal, with a time-averaged amplitude of $\simeq$1\%\ (half amplitude).
The highly coherent pulsation points towards a rapidly rotating neutron star 
to be present in 4U\,1636$-$53; it further supports the
connection between burst-oscillation frequencies and the neutron-star spin frequencies 
(see, e.g., \cite{SB2003}, and references therein). 

So far, the superbursts have only been observed in
sources with persistent pre-burst luminosities, $L_{\rm pers}$, of $\sim$0.1--0.25 times
the Eddington luminosity, $L_{\rm Edd}$ \cite{W2001,Ketal2002a} 
(see Table~\ref{properties}); apparently, the underlying mass accretion rates create 
ideal circumstances for the origin of the superbursts \cite{CB2001}.

It was already clear from the {\it Beppo\-SAX}/WFC observations that the superburst
affects the normal X-ray burst activity. {\it No} normal X-ray bursts were found,
during the continuous monitoring observations of 4U\,1735$-$444, for about 7.5 days immediately 
after the superburst, despite the X-ray flux level being similar to
occasions when the source did exhibit normal X-ray bursts.
This became even more apparent when analysing the {\it Beppo\-SAX}/WFC observations
of KS\,1731$-$260 and Ser\,X-1 (Fig.~\ref{ser_ks}). These observations revealed that before 
the superburst the source
was happily showing normal X-ray bursts, then for about a month after the superburst the normal X-ray bursting
ceased, and finally it resumed bursting again afterwards. For 4U\,1636$-$536 a similar
cessation time scale can be inferred (see Table~\ref{properties}).

From the previous paragraphs it may have become clear that the properties of the superburst observed from
4U\,1820$-$303 seem to be different from those of the other superbursts: 
the peak temperature and peak flux reached
during the superburst were higher with respect to other superbursts, 
which resulted in a somewhat larger fluence
compared to the other superbursts. Also, the peak flux of the precursor burst was somewhat
weaker than that of the superburst. These are related to the fact that the superburst of 4U\,1820$-$30 
showed strong radius expansion, whereas the other superburst did not show evidence for such a phase.
Note also, that the peak of the superburst is reached much faster (by about a factor 10) than seen
in 4U\,1636$-$536 and KS\,1731$-$260.

4U\,1820$-$303 being an exception to the `superburst rule' might be related to the fact that its
orbital period is much smaller than those of 4U\,1636$-$536, 4U\,1254$-$690
and 4U\,1735$-$444 (see above). Systems like 4U\,1820$-$30 are thought to contain a
degenerate He-donor (e.g., \cite{Retal1987}), whereas for the others it has been shown that the donors
provide a mix of H/He (presumably solar; \cite{Metal1978,Aetal1998}).
This is consistent with 4U\,1820$-$30 showing only He-flashes, whereas the other sources clearly
have shown X-ray bursts due to unstable mixed H/He burning (except GX\,3+1, see
\cite{Hetal2003}). 

\section{Some theoretical interpretation}

Generally, the superbursts last too long and their energy release is too much in order to explain them
through unstable burning of H and/or He (see, e.g., \cite{SB2002,Setal2003a,NH2003}). 
Moreover, regular normal X-ray bursts are seen up to the occurrence
of the superburst (e.g., \cite{Ketal2002a}), including the precursor burst. 
The long rise and decay times of the superbursts are consistent with 
unstable burning from a greater depth, i.e., below the H and/or He layer.
It has, therefore, been suggested that unstable burning of C 
is the origin of the superbursts \cite{CB2001,SB2002} (see also Cumming, this volume).

\begin{figure}[ht]
\resizebox{7.5cm}{!}{\includegraphics[clip, bb=92 270 480 737]{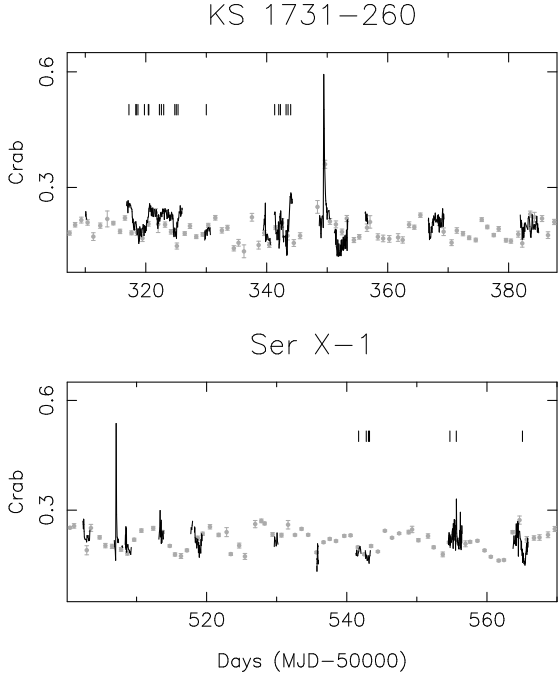}}
\caption{Long term light curves of KS\,1731$-$260 ({\bf top} \cite{Ketal2002a}) and 
Ser\,X-1 ({\bf bottom} \cite{Cetal2002}) around
the time of their superburst. The continuous lines indicate the {\it Beppo\-SAX}/WFC observations
at a time resolution of 5\,min,
the grey data points are the daily-averaged {\it RXTE}/ASM dwells, the vertical bars indicate
the time of occurrence of a normal type~I X-ray burst. Note that type~I X-ray bursting ceases
for about a month after the superburst.}
\label{ser_ks}
\end{figure}

If the accreted material onto the neutron star is pure He, C can be produced when
He is burned stably or unstably (nearly 100\%\ and $\sim$3\%, respectively \cite{SB2002,Wetal2003}). 
This applies to the He accretor 4U\,1820$-$303 which 
shows long periods of high intensity during which no bursts occur, consistent with a 
period of stably burning He. Note that unstable C burning can only reproduce the superburst observed
when taking into account neutrino losses and significant heat flux deeper into the neutron star
\cite{SB2002}.
Recurrence times on the order of 1--2~years \cite{C2003} are expected
(but see \cite{SB2002} who quote a recurrence time of about 10~years). 
However, if the (degenerate) donor still provides some H, the recurrence times may 
be 5--10~years \cite{C2003}. 

If the accreted material onto the neutron star is a mixture of H and He, C can also be produced
by either unstable or stable burning of H/He, but only in rather limited amounts
($\lesssim$1\%\ and $\sim$10\%, respectively \cite{Wetal2003,Setal2003b}).
In the months before the superburst and/or after the normal X-ray burst cessation period,
normal X-ray bursts in the H-rich accretors occur irregularly with a mean rate of about 3 per day
\cite{Cetal2002,Ketal2002a} (see, e.g., Fig.~\ref{ser_ks}; see also
\cite{Cetal2003}). This indicates that at least some of the accreted material
is burning stably around the time of superburst. This may suggest
that superbursts only occur in systems where in between normal X-ray bursts 
stable burning takes place (see also \cite{Zetal2003}; Cumming, this volume). In this respect it
is interesting to make notice of other frequently bursting X-ray sources in the 
Galactic Center region, such as 4U\,1702$-$429 and A1742$-$294, which have similar normal X-ray burst
occurrence times \cite{Cetal2003}. These may be good candidates for
exhibiting superbursts as well.

Cumming \&\ Bildsten \cite{CB2001} have shown that it is possible to ignite small amounts of C
for the H/He accretors, when
it resides in a bath of heavy elements. These heavy elements are the products of the 
unstable burning through the rp-process during the mixed H/He X-ray bursts.
In this case the superburst recurrence times depend on the accretion rates onto the neutron
star, being in the order of decades, a year to a decade, or a week to a month,  
for accretion rates of about 0.1, 0.3, or 1 times the Eddington
accretion rate, respectively. 
More recently, it was found that due to the high temperatures reached during the superburst,
a photo-disintegration runaway may be triggered. With this mechanism the heavy elements
are converted into iron group elements. This gives rise to an energy production
which is comparable to the C burst itself \cite{Setal2003a} (but see \cite{Wetal2003}). 
Since the X-ray flux doubles immediately after the precursor burst of, e.g., 4U\,1636$-$536, it is interesting to
speculate that the precursor burst may have triggered this photo-disintegration process,
and that unstable C burning had already started before that (which may have 
triggered in turn the precursor burst, see, e.g, \cite{SB2002}).
Note that if the precursors are due to ignition of the He layer by flux from the
C burning, then the mass of the He layer will be some amount less
than the critical mass needed for igniting a normal X-ray burst. Therefore,
it is reasonable for the precursors to be weak compared to a normal X-ray burst
(Cumming, this volume).

Another scenario was proposed by Kuul\-kers et al.\ \cite{Ketal2002a}, who suggested that H left over from 
the burning of the H/He layer is reignited by electron capture,
with subsequent capture of the resulting neutrons by heavy nuclei, deeper into the neutron star
(i.e., in the same bath as mentioned above).
In this case relatively large amounts of H have to be left over in order to satisfy the energy release.
Recent calculations have shown, however, that H is more or less depleted after the H/He burning
\cite{Setal2001,Wetal2003}, making this scenario less viable.
Nevertheless, recurrence times on the order of a year or less are to be expected 
\cite{Ketal2002a,K2002}.

The bottomline here is, that at present one can not strongly rule out the proposed models, purely based
on the recurrence times. For that one needs more stringent time scales from multiple superbursts
in a source.

\section{Conclusion}

The recent discovery of eight long X-ray flares, superbursts, seen in seven X-ray burst sources 
share many of the characteristics of type~I X-ray bursts.  What distinguishes them
from type~I X-ray bursts are the long duration (exponential decay
times of a few hours), the large fluences ($\sim$10$^{42}$\,erg),
and the extreme rarity.  
They are therefore attributed to a new mode of thermo-nuclear runaway events.
The current view is that the superbursts are caused by the unstable
burning of the ashes of the (un)stable H and/or He burning.
Such bursts in principle thus not only tell us about properties of material buried below the
H and/or He layer, but also about the burning of the H and/or He layer
itself (see, e.g., \cite{C2003,Wetal2003}; Cumming, this volume). 

With monitoring programs on satellites currently operating ({\it Integral}, {\it RXTE}) 
as well as future missions (e.g., {\it Swift}), one hopes to discover more of these
powerful events. On the other hand, a scan through archival data could reveal other
(parts of) superbursts.
Multiple superbursts from the same source may help to constrain their recurrence times,
whereas the study of superbursts from other sources may help to understand the 
environment in which the superbursts reside. Crucial information comes also
from the type~I X-ray burst behaviour months to years before and after a
superburst. Dedicated programs, such as to continuously monitor the Galactic Center region
with a wide field of view (e.g., {\it MIRAX}), are ideal for such studies.

\vspace{-0.3cm}
\section*{Acknowledgements}
\vspace{-0.2cm}
The {\it RXTE}/ASM data are provided by the ASM/RXTE teams at MIT and at the RXTE SOF and GOF at NASA's
GSFC. Andrew Cumming and Jean in 't Zand are acknowledged for providing comments on earlier versions
of the manuscript. I thank Joeri van Leeuwen for drawing my attention to Superman.

\newpage

\section*{Epilogue}
\label{epilogue}

Although the word `superburst' was first used by Wijnands \cite{W2001} to describe the 
powerful, very long X-ray flares, historically it should be noted
that a relatively strong type~I X-ray burst seen from 4U\,1728$-$34 was denoted with the same name.
I quote:
{\it There is one burst (we call it the `super burst') which is about 3 times 
more energetic than the average burst from} [4U\,1728$-$34] \cite{Betal1984}.
Interestingly, scientists in a completely other research area struggled with a similar nomenclature
`problem'. This was related to exceptional phenomena seen from the Steamboat Geyser in 
Yellowstone National Park. I quote:
{\it ... the power of the steam phase was frequently a mind-numbing sensory overload.
The most common reaction of the most experienced observer was:
`I don't believe it!' ...  At first the authors called this unusual display an
`Oh my god' burst. Later that evening while describing the
unusually powerful event to other observers, it was called
a superburst for want of a better name. The
authors regret that this unoriginal term has become the accepted
name for the phenomena, since the same term has been in common
usage for years to describe unusually powerful eruptions of Great
Fountain Geyser} (\cite{Setal1989}, and references therein).

I further quote:
{\it Great Fountain Geyser {\rm (Fig.~\ref{greatfountain})} is a fountain-type geyser. 
The interval between
eruptions ranges from 9 to 15 hours, but its short term average interval is usually 
stable enough that the eruptions can be predicted to within an hour or two. 
Great Fountain Geyser's maximum height ranges from
about 75 feet to over 220 feet. The duration of an eruption is usually about one hour, 
but durations of over two hours have been seen.

\begin{figure}[ht]
\begin{center}
\resizebox{5.6cm}{!}{\includegraphics[clip, bb=168 225 426 617]{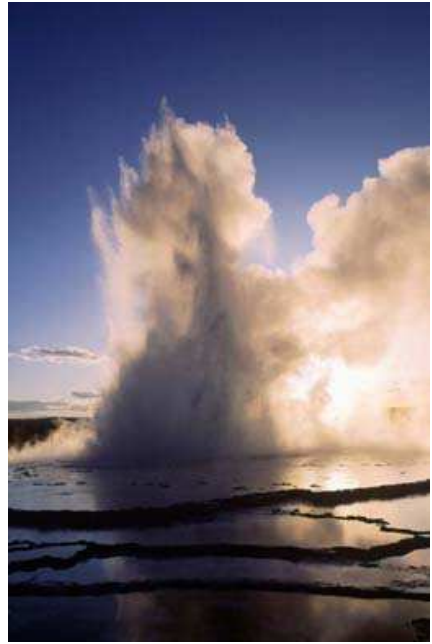}}
\caption{An eruption of the Great Fountain Geyser. This geyser can be found in the Lower Geyser Basin, 
Yellowstone National Park.}
\label{greatfountain}
\end{center}
\end{figure}

Great Fountain Geyser has two types of truly spectacular behaviour. 
A superburst is an exceptionally tall burst of water, over 150 feet.
Some superbursts have reached 230 feet. Superbursts, when they occur, 
are usually the first burst of the eruption, but they have
been known to sometimes occur later in the eruption. 
A {\rm blue bubble} occurs when a calm and still pool of water is domed up by
a large expanding steam bubble. As the steam bubble rises and expands, 
the entire 16 feet wide pool of water is lifted and domed outward creating a beautiful 
{\rm blue bubble}. Once the steam reaches the surface, the water explodes outward and upward.
{\rm Blue bubbles} most commonly occur at the start of the eruption, but they have been known 
to occur at the start of other active periods. 
A fair number of {\rm blue bubbles} result in a superburst, but not all. 

Great Fountain Geyser sometimes goes through a {\rm Wild Phase}. During a {\rm wild phase}
the geyser seems to forget how to end an eruption. A 10 to 50 feet play continues for 
hours to days. Once the play finally ends, Great Fountain Geyser usually takes a few days
to recover before returning to ``normal'' eruptions. Interestingly, {\rm wild phases} mainly occur 
late in the year.} 
(Courtesy `The Geyser Observation and Study Association';
for more information about the Great Fountain Geyser and other geysers 
I refer to {\tt http://www.geyserstudy.org}.)

\end{document}